%% file: make_all.tex
\documentclass{he_symp}
\usepackage{psfig,graphicx,epsfig}
\usepackage{color}
\usepackage{amsmath,amssymb,epic,eepic,array}
\begin{document}
\renewcommand{\FirstPageOfPaper }{ 44}\renewcommand{\LastPageOfPaper }{ 53}\include{he_symp_shearer}
\clearpage

\end{document}

%% file: he_symp_shearer.tex

\title{Why study pulsars optically?}
\author{A. Shearer\inst{1}, A. Golden\inst{1}}  
\institute{Computational Astrophysical Laboratory, National University of Ireland, Galway, Ireland}
\maketitle

\begin{abstract}

Observations of the five confirmed optical pulsars indicate that the peak
emission scales according to the magnetic field strength at the light
cylinder. To the accuracy that such low number allows we show that
this gives gives further confirmation that a straightforward
synchrotron model such as \cite{pac87} still has validity.  The derived
relationships indicates that the emission mechanism is common across
all of the pulsars and towards the edge of the light cylinder. In the
future observations should include near and far infra red work to
determine the long wave self absorption cut-off and polarization
observations of all pulsars to restrict (to first order) emission zone
geometry \\

\end{abstract}

\section{Introduction}

Since the first optical observations of the Crab pulsar in the late
1960s (\cite{coc69} only four more pulsars have been seen to pulsate
optically (\relax  PSR B0540-69; \sloppy \cite{mid85}; Vela: \cite{wall77};
B0656+14; \cite{shear97}; Geminga; \cite{shear98}).
Four of these five pulsars are probably too distant to have any
detectable optical thermal emission using currently available
technologies. For the fifth (and faintest) pulsar, PSR 0633+17,
spectroscopic studies have shown the emission to be predominantly
non-thermal (\cite{mar98}). With these objects we are therefore
seeing non-thermal emission, presumably from interactions between the
pulsar's magnetic field and a stream of charged particles from the
neutron star's surface. Many suggestions have been made concerning the
optical emission process for these young and middle-aged pulsars.
Despite many years of detailed theoretical studies and more recently
limited numerical simulations, no convincing models have been derived
which explain all of the high energy properties. There are similar
problems in the radio but as  the emission mechanism is radically
different (being coherent) only the high energy emission will be
considered here. However, the most successful analytical model
(both in terms of its simplicity and longevity) has been proposed by
\cite{pac71} and in modified form by Pacini and Salvati (1983; PS83, 1987;
PS87). In general the model proposed that the high energy emission
comes from relativistic electrons radiating via synchrotron processes
in the outer regions of the magnetosphere, and that the location and
approximate dimensions of this emission region scales with the
magnetic field in the vicinity of the light cylinder. In this paper we
examine the validity of their approach and show that it adequately
explains the observed phenomena.
               
In recent years a number of groups have carried out detailed
simulations of the various high-energy emission processes. These
models divide into two broad groups - between acceleration and
emission low in the magnetosphere (Polar Cap models \cite{dau96}) and
those with the acceleration nearer to the light cylinder (Outer-Gap
models \cite{che00} and refs therein). Both models have problems explaining the observed
features of the limited selection of high energy emitters. However
both models suffer from arbitrary assumptions in terms of the
sustainability of the outer-gap and the orientation of the pulsar's
magnetic field to both the observers line of sight and the rotation
axis. Furthermore some observational evidence, see for example
\cite{eik97}, severely limits the applicability of the outer-gap to
the emission from the Crab pulsar. However they have their successes -
the total polar-cap emission can be understood in terms of the
Goldreich and Julian current from in or around the cap; the Crab
polarization sweep is accurately produced by an outer-gap variant
\cite{rom95}. Until we have more detailed observations - and in the
optical all aspects of the radiation can be measured (intensity,
timing, energy and polarization) - on more objects the situation will
not improve.
               
It is the failure of the detailed models to explain the high energy
emission that has prompted this work. We have taken a phenomenological
approach to test whether Pacini type scaling is still applicable. Our
approach has been to try and restrict the effects of geometry by
taking the peak luminosity as a scaling parameter rather than the
total luminosity. In this regard we are removing the duty cycle term
from PS87. It is our opinion that to first order the peak emission
represents the local power density along the observer's line of site
and hence reflects more accurately the emission process within the
pulsar's magnetosphere. Previous work in this area (see for example
\cite{gold95}) looked at the total efficiency, spectral index and found
no reasonable correlation with standard pulsar parameters - age,
period and spin down rate. Their work was hampered by not including
geometry and being restricted to the then three known pulsed
emitters. Since then the number of optical pulsars has increased to
five. However we are still dealing with very weak statistics
in all cases bar that of the Crab pulsar.

\section{Current Observations - An Overview}

\subsection{Introduction}

The three brightest pulsars (Crab, Vela and PSR B0540-69) are also
amongst the youngest. Tables 1 and 2 show the current observed
limitations of optical pulsars. Only the Crab is sufficiently bright
for individual pulse work to be performed. For the faintest objects
very high throughput instruments (including polarimetry) are needed to
do more than simple photometry.

Table 4 shows the basic parameters for these
objects. However all these pulsars have very different pulse shapes
resulting in a very different ratio between the integrated flux and
the peak flux. Table 4 also shows this peak emission (taken as the
emission in 95\% - 95\% portion of the largest peak). Their distances
imply that the thermal emission should be low (in all cases $<$ 1\% of
the observed emission). One observational note the Crab is by far the
brightest optical by a factor of 250:1. Only the Crab is bright enough
for individual pulse work - of all Stokes' parameters. Furthermore it
is the only one for which Stokes' parameters can be accurately measured
throughout the pulsar's rotation period.

\begin{table}
\caption{Observed fluxes from the observed optical pulsars in B and 
scaled to the VLT. Note the sky background is of the order 1300 
photons/sec/arcsec$^2$ for the VLT.}

\begin{small}

\begin{tabular}{lcc}

Name & Int.  Flux & Int. Flux \\ & ($m_B$) & ph/sec/VLT \\

Crab                  & 17      & 90,000  \\
PSR 0540-69           & 23       & 370  \\
Vela                  & 24       & 150  \\
Geminga               & 25.5     & 37 \\
PSR 0656+14           & 26       & 23 \\
(PSR 1509-58          & 25.7     & 30) \\

\end{tabular}
\end{small}
\end{table}

\begin{table}

\caption{Observed fluxes from the observed thermally emitting pulsars in B and scaled to the VLT. Note the sky background is of the order 1300 photons/sec/arcsec$^2$ for the VLT.}

\begin{small}

\begin{tabular}{lcc}

Name & Int.  Flux & Int. Flux \\ & ($m_B$) & ph/sec/VLT \\

PSR 1929+10	&	$>$26.2 	& $<$20 \\
PSR 0950+08	&	27.2    &  7 \\
RXJ 1856.5-37.54 &	 25.5	&  36 \\
PSR 0656+14  	&	$>$26.8	&  $<$14 \\
Geminga 	&	$>$27     &  $<$10 \\
PSR B1055-52	&	24.9 (U) &  $\approx$50 \\

\end{tabular}
\end{small}
\end{table}

\subsection{The Gold Standard - the Crab pulsar}
       
The Crab pulsar's phase plot is well known - see Figure 1 for R/I
(6000-8500 \AA) measurements taken with an avalanche photo-diode (APD) on
the WHT over 3 nights of observation. Figure 2 shows the peak region
indicating the very small plateau. Figure 3 shows 2-d time resolved
V band MAMA data - in particular we note emission during the  
`off pulse' region (phase 0.77-0.84). This is discussed
in more detail below.

\begin{figure}
\centerline{\psfig{file=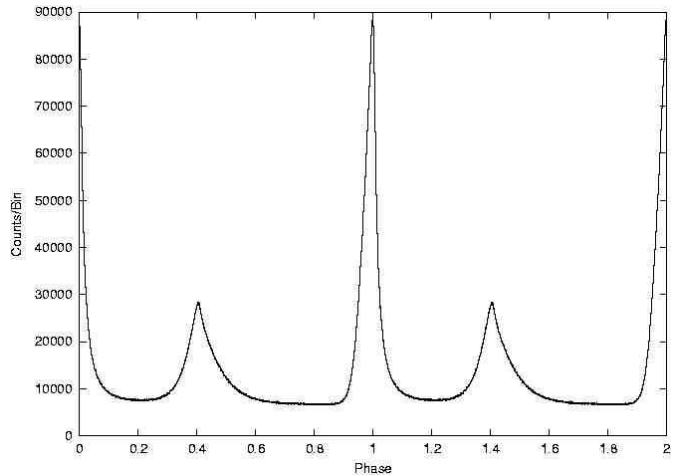,angle=-90,width=8.8cm,clip=} }
\caption{Crab pulse profile in R and I with 3000 phase bins taken in November 1999 using an APD in the 
TRIFFID photometer}
\end{figure}

\begin{figure}
\centerline{\psfig{file=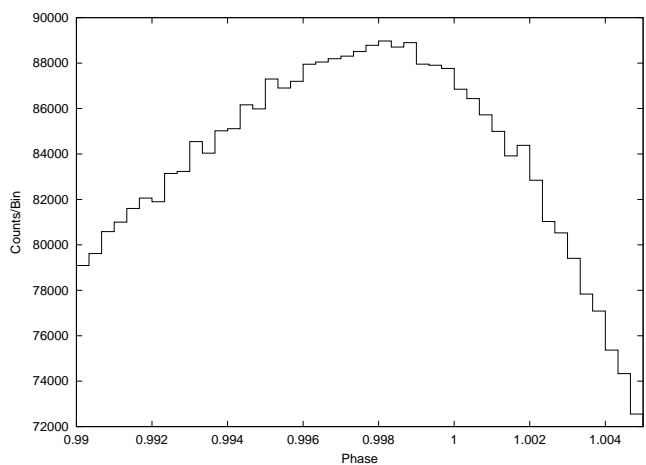,angle=-90,width=8.8cm,clip=} }
\caption{The peak of the Crab pulse (from the same data set as Figure 1 showing the small or 
non-existent plateau}
\end{figure}

\begin{figure}
\centerline{\psfig{file=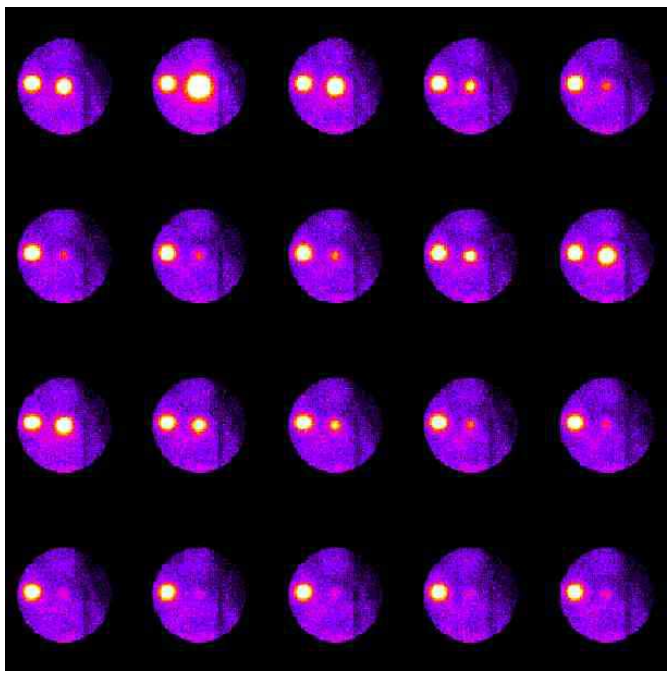,width=8.8cm,clip=} }
\caption{Phase resolved images taken over three nights using a 2-d MAMA camera in B on the Russian 6m telescope.}
\end{figure}

Over the last thirty years the Crab pulsar has been studied more
extensively than any other. 
It is unique in that its age is known accurately, 
it is
relatively close and bright. One of the main predictions of
\cite{pac71} was the prediction that the Crab luminosity should reduce
by $\approx$ 0.005 mags/year. \cite{nas96}'s estimate of $0.008 \pm
0.004 $/ year is in agreement with this (see Figure 4). However more
detailed observations are required - preferably with the same
telescope, detector and filter arrangements to conform this and
remove systematic errors.

\begin{figure}
\centerline{\psfig{file=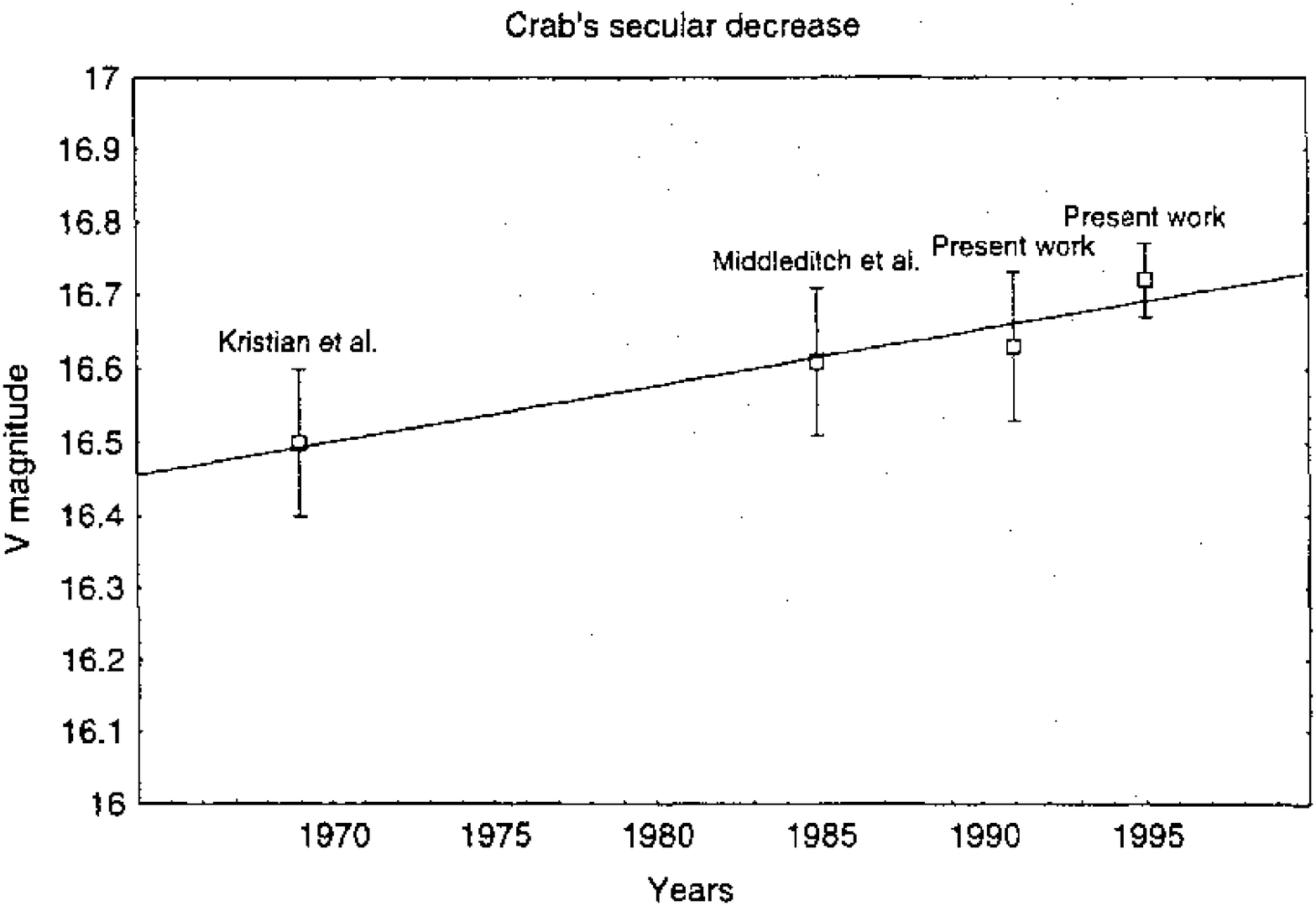,width=8.8cm,clip=} }
\caption{Secular change on Crab luminosity - from \cite{nas96}}
\end{figure}

Smith et al.~(1988) made detailed 1-d optical measurements of the phase resolved 
polarization, displayed in Figures 6 \& 7. They noted that the
polarization percentage followed the same shape over the main and
interpulse. However interestingly they noted that the highest level of
polarization was evident in the `off'-pulse region. \cite{smith88}
went to some considerable lengths to remove the background nebular
polarization component, and thus viewed this latter result as a
consequence of the `unpulsed' component being a cusp region between
the two pulses. They also noted that the polarization angle swings in
similar way through the main and interpulse which they interpreted as
indicating a similar source (and geometry) of the radiation for both
pulses, but coming from opposite poles. A similar, albeit slightly
phase shifted, result has been observed by
\cite{rom01} using the new Transition Edge Detector (see Figure 18).

The Crab's spectrum shows no evidence for a low energy break (see for example
\cite{sol02} and references therein) towards the IR regime, rather some
suggestion of a `levelling' of the spectral index (see Figure 8). Thus
far no other optical pulsar has been seen to show evidence for a
spectral break at low energies, although we are to some extent
constrained by their faintness and the limitations of current high
speed detector technology in this waveband.  A definitive measurement
of the Crab suggested `roll-over' - indicative of some form of
optically thick transition such as self-absorption - in the IR is one
of the remaining crucial measurements which should be performed in the
next few years.

\begin{figure}
\centerline{\psfig{file=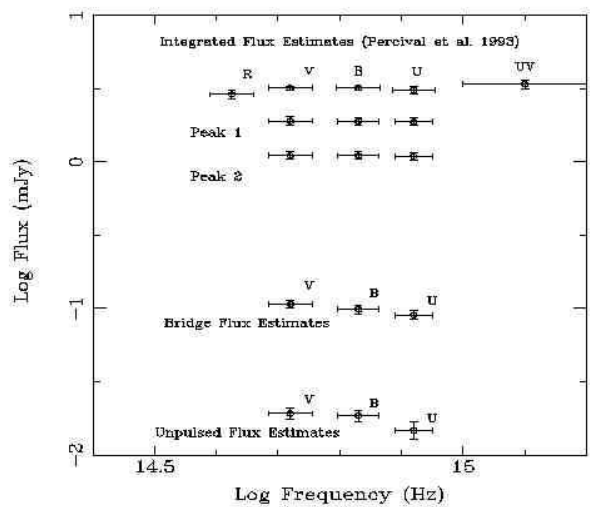,width=8cm,clip=} }
\caption{$UBV$ spectrophotometry of the Crab pulsar's light curve components
\cite{gold00}}
\end{figure}

An unpulsed optical component of emission has been noted for some
time, since the work of \cite{pet78}, although accurate high speed 2-d
photometry as always dogged previous efforts to flux the apparent
emission. \cite{gold00} using a MAMA based system obtained sufficient
S/N and critically, excellent background statistics to allow for an
accurate determination of the flux of this component - clearly apparent in
Figure 3, and broad-band spectra in UBV in Figure 5. This unpulsed region,
described by \cite{gold00}, can be interpreted in two ways - either a
chance alignment of (e.g. a knot) or as a near constant background
component to the emission. The star like point source unpulsed image
in Figure 3 showed no significant positional shift compared to the
integrated position - a shift of at least 3 pixels would be expected
if was associated with the red knot (0''.65 from the pulsar) (see
\cite{hes95} and \cite{sol02}).  Furthermore the intensity of the knot
given by \cite{sol02}, does not explain the R/I background level shown in
Figure 10.

\begin{figure}
\centerline{\psfig{file=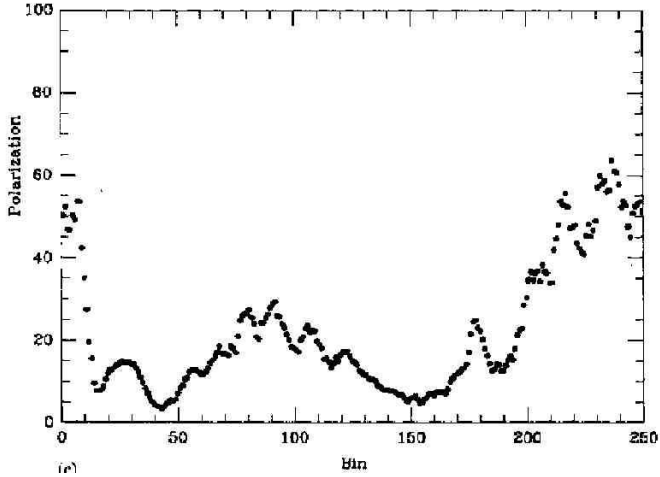,width=8.8cm,clip=} }
\caption{Crab polarization profile from Smith et al 1988}
\end{figure}

\begin{figure}
\centerline{\psfig{file=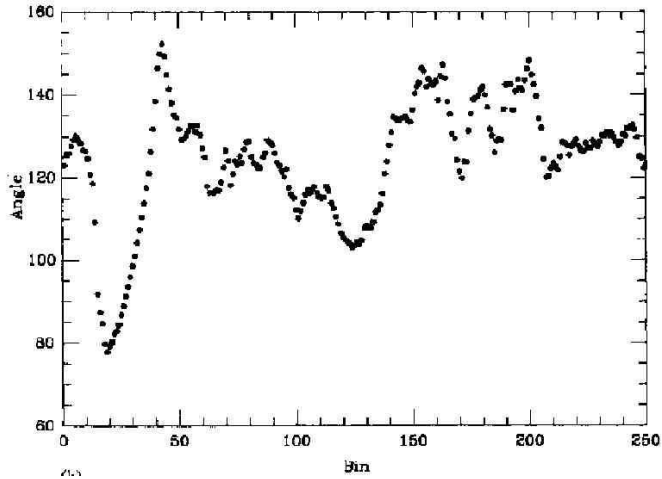,width=8.8cm,clip=} }
\caption{Crab polarization angle from Smith et al 1988}
\end{figure}

\begin{figure}
\centerline{\psfig{file=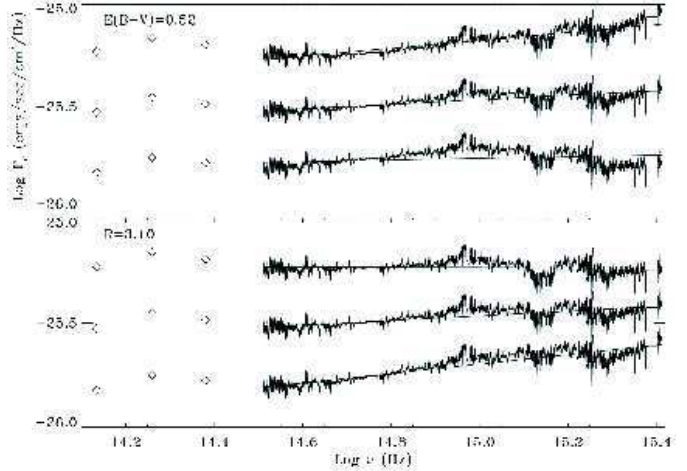,height=6.3cm,clip=} }
\caption{Crab pulsar spectrum \cite{sol00} with differing values of
E(B-V) and R. For E(B-V) = 0.52, fits are [R = 2.9 (top), 3.1 \& 3.3],
for R = 3.1, fits are [E(B-V) = 0.49 (top), 0.52 \& 0.55].
Spectra have been shifted by -0.3, 0.0 \& +0.3 dex in `flux' for
clarity. \cite{eik97} $JHK$ data are included - these are
consistent with 2MASS photometry for the pulsar.}
\end{figure}

\begin{figure}

\centerline{\psfig{file=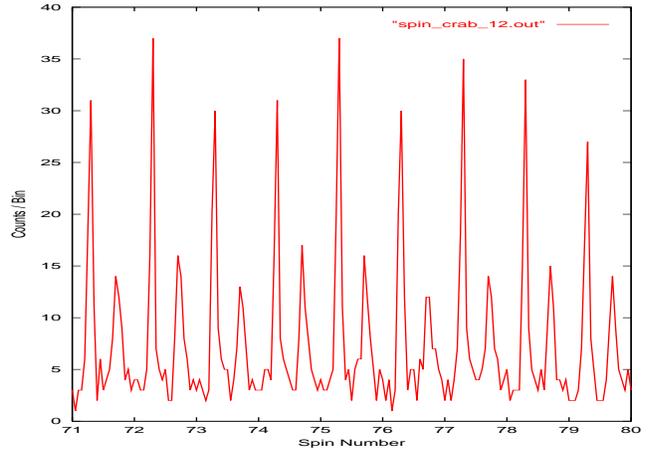,width=8.8cm,height=6cm,clip=} }

\caption{Individual Crab pulses - statistical analysis shows no deviation from the expectation of Poissonian statistics. There is no observed correlation with the radio giant pulses \cite{shear02b}}

\end{figure}

\begin{figure}
\centerline{\psfig{file=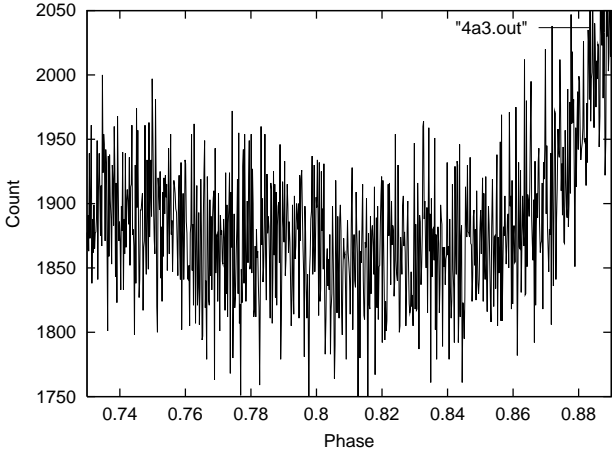,width=8.8cm,clip=} }

\caption{'Off' pulse region of Crab pulsar - taken from the full pulse in figure 1. The sky background in this image is at 4390 counts. The knot, see for example \cite{sol00} \& \cite{sol02}, situated 0''.6 from the pulsar (and within our aperture) would contribute a further 650 counts/bin}

\end{figure}

\subsection{The Remaining Optical Pulsars}
       
Table 1 gives a summary of the observational data of the 5 pulsars
seen to give pulsed magnetospheric emission. Of all the optical
pulsars PSR B0633+17 (or Geminga) is perhaps the most
controversial. Early observations (\cite{hal88}; \cite{big88})
indicated that Geminga was an $\approx$ 25.5 m$_V$ object. Subsequent
observations including HST photometry appeared to support a thermal
origin for the optical emission combined with a cyclotron resonance
feature (\cite{mig98}).  The high-speed optical observations of
\cite{shear98} combined with spectroscopic observations (\cite{mar98})
contradict this view.  Figure 11 shows the broadband data plotted on
top of the Martin et al Keck spectra, we have also included the pulsed B
point acquired via a ground-based MAMA system (\cite{shear98}) and recent 
Subaru data (\cite{kaw02}). This
combined dataset indicates a fairly steep spectrum (with spectral
index of $\sim$ 1.9) consistent with magnetospheric emission. It was on the
basis of these results that
\cite{gold99} were able to give an upper limit of R$_\infty$ of about
10km by considering the upper limits to the unpulsed fraction of the
optical emission from Geminga as an upper limit for the thermal
emission.  The evolving picture for Geminga is that the soft X-ray and
EUV data is predominantly thermal with a magnetospheric component
becoming dominant at about 3500 \AA.

\begin{figure}
\centerline{\psfig{file=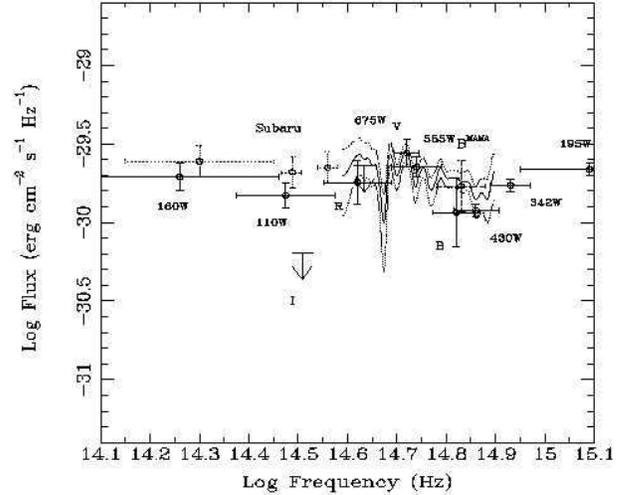,width=8cm,clip=} }

\caption{Spectra and Photometry of Geminga. The spectrum (\cite{mar98} and the
agreement between it and the integrated photometry (\cite{mig98}). B
is the pulsed flux from \cite{shear98}. The HST points are from
Pavlov, private communication - these are in close agreement with the
IR points presented at this workshop \cite{kaw02}.  }
 
\end{figure}
                
As regards PSR B0656+14, the pulsar is generally agreed to be
predominantly a non-thermal emitter in the optical, becoming
a thermal emitter at wavelengths shorter than about 3000
\AA \cite{pav97}. There is a discrepancy between the radio distance
based upon the dispersion measure and the best fits to the X-ray
data. From radio dispersion measure a distance of $ 760 \pm 190 pc $
can be derived at odds with the X-ray distance of $250-280 pc$ from
$N_H$ galactic models. A similar low estimate for the distance was
given by \cite{kap01}. The optical spectrum is best fitted by a two
component model with the power law component being significantly
steeper than the Crab. Both Geminga and PSR B0656+14 have steep
spectra compared their younger cousins. The observed light curve is
roughly sinusoidal presumably indicative of closer alignment between
the observer and the rotation axis than for the Crab and Geminga.

PSR B0540-69, the second brightest optical pulsar, is also the most
distant. It is regarded as a twin of the Crab as it is of similar age,
spin-down energy, breaking index and it is embedded within its own
plerionic remnant. It has a broad light-curve with two distinct peaks
- see figure 12. The difference between its light curve and that of
the Crab is likely due to geometry. This pulsar has also had its braking
index measured, $n=2.2 \pm 0.02$ (\cite{boy95}). This (like the Crab)
differs from the pure dipole value of 3. \cite{mid87} measured no residual polarization from this pulsar - albeit at low
significance.

\begin{figure}
\centerline{\psfig{file=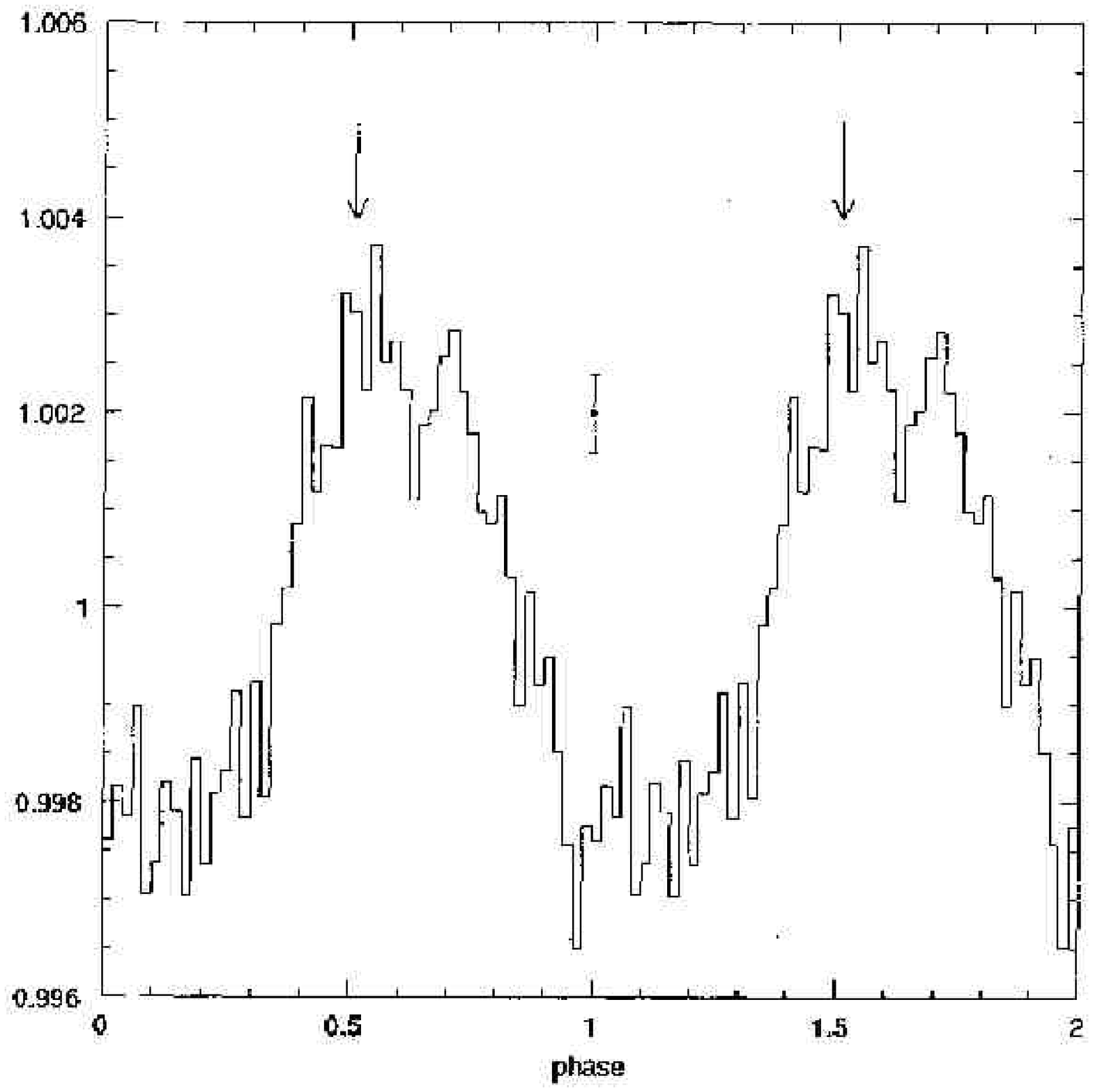,width=8.8cm,clip=} }

\caption{Light Curve of PSR B0545-69 from \cite{gou92}}

\end{figure}

The Vela pulsar has not been studied extensively in the optical. The
best light curve from \cite{gou96} shows strong similarity between the
$\gamma$-ray profile and the optical. More data is required to
establish both the detailed structure of the light curve and the
pulsar's polarization properties.

\begin{figure}
\centerline{\psfig{file=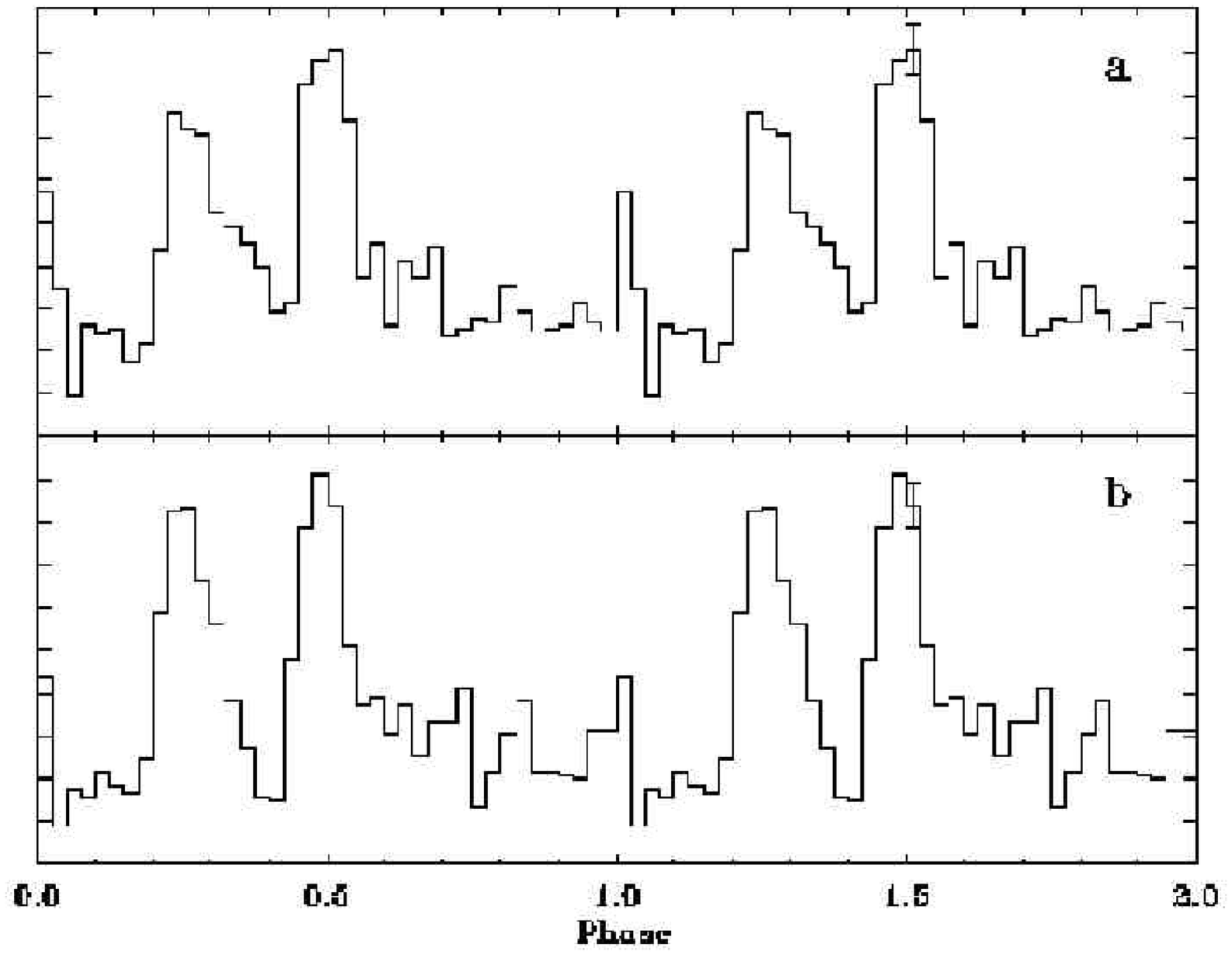,width=8.8cm,clip=} }
\caption{Vela light curve from \cite{gou96}}

\end{figure}

\subsection{Discussion}

In Table 1 we have indicated the peak flux normalized to the Crab
pulsar. When comparing the individual pulsars (which will have a range
of viewing angles and differing magnetic/rotation axes), we have to
account for the viewing angle (related to the pulse duty cycle and
separation in terms of phase) as well as the total flux. Furthermore for each source, we
have to account for the shape of the pulsar's spectrum. 
PS87 indicates for emission above the
low energy cut-off that the ratio of the fluxes from two different
pulsars can be given by Equation 1 if one ignores the effect of duty
cycle and pitch angle (the suffixes refer to each pulsar).  Here
$F_{\nu,n}$ refers to the flux at the observed frequency $\nu$ for
pulsar n (similarly for the magnetic field strength, B, and period,
P). The observed energy spectrum exponent is given by $\alpha_n$. The
duty cycle can be accounted for by only considering the peak
emission. The pitch angle being beyond the scope of this work and
assumed to first order to be invariant. Equation 2 shows the same
formulation for the outer field case.  \\

\begin{equation}
           \frac{F_{\nu,2}}{F_{\nu,1}} \propto (\frac{\nu_{1,0}}{\nu})^{\alpha_2-\alpha_1} 
(\frac{B_{2,0}}{B_{1,0}})^{4-\alpha _2} (\frac{P_{2}}{P_{1}})^{3\alpha-9} 
\end{equation}

Scaling to the transverse field would give:

\begin{equation}
           \frac{F_{\nu,2}}{F_{\nu,1}} \propto (\frac{\nu_{1,0}}{\nu})^{\alpha_2-\alpha_1} 
(\frac{B_{2,0}}{B_{1,0}})^{4-\alpha _2} (\frac{P_{2}}{P_{1}})^{3} 
\end{equation}

Given the observed peak luminosities we investigated the correlation
between the peak emission and the surface field and the tangential
light cylinder field. Figure 14 shows the relationship between the
peak luminosity with the outer (tangentially at the light cylinder)
magnetic field strength, $B_T$, Goldreich-Julian current and canonical
age ($\frac{P}{2\dot{P}}$). A clear correlation is seen with all
these parameters. We are interested in investigating the implications
of a correlation between the peak luminosity and the transverse
field. We accept the the strong correlation with G-J current would
under pin both emission from polar as well as outer regions.

A regression of the form:\\

\noindent
Peak Luminosity $\propto B_T^{\beta}$ \\

\noindent
was performed for the empirical peak luminosity this leading to a
relationship of the form :-\\ 

\noindent
Peak Luminosity $\propto B_T^{2.86 \pm 0.12 }$ \\ 

\noindent
significant at the 99.5\% level and consistent with the high average
exponent shown in Table 1.  Figure 14 shows the predicted peak
luminosity from Equation 2 against our observed peaks accounting for
the differing observed energy spectrum exponent at 4500 \AA.  The
slope is 0.95 $\pm$ 0.04 and significant at the 99 \% level.  Whilst
informative it still goes no further than previous attempts to
understand the phenomena of optical emission. Importantly, we note
that the flattening of the peak luminosity relationship for the older,
slower pulsars is consistent with their having a significantly steeper
energy spectrum than the younger pulsars, see Table 2. Whilst this is
consistent with the emission zone being optically thick we consider it
to be more likely that it reflects a larger emission region for these
pulsars rather than the younger ones.  This broadly concurs with PS87
argument of $\Delta$ (the size of the emission zone) scaling with P.

\begin{figure}
\centerline{\psfig{file=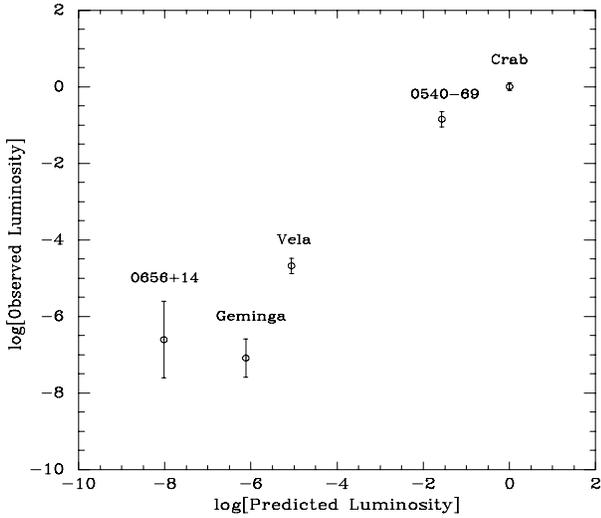,width=8cm,clip=} }
\caption{Observed Peak optical luminosity vs predicted} 
\end{figure}

The polarization of the Crab's `off-region' can be interpreted as
coming from a different source to the main pulsed emission. Smith et
al's argument that the off emission represented a cusp between the two
main pulses relies upon the `off'-pulse being a genuine cusp and not
flat as observed by \cite{shear02a}. The flat region - see Figure 10 -
covers the peak of polarization noted by Smith et al. Another
explanation is that the net polarization is a combination of two
components; the remnants of a pulsed `normal' magnetospheric component
and a secondary unpulsed `off-pulse' component. The derived
polarization profile is then just the intensity normalized combination
of these two processes. As the bridge component has twice the
intensity of the `off'-pulse we would expect the degree of
polarization to be half that of the off-pulse region - as
observed. The swing in polarization can therefore be understood in
terms of the unpulsed component becoming dominated by by the pulsed
component during the rising edge of each pulse. The relative
contributions being due to normal quadratic addition of the Stokes
parameters - i.e. dominated by the relative intensity. Figure 17 shows
the derived polarization profile if it only comes from a steady low-level
background source. The shape agrees well with the previously measured
profile from Smith et al, except in the region around the main and
inter-pulse - where the main pulsed polarization component will occur.

It seems clear that from both polarization studies (\cite{smith88};
\cite{rom95}) and from this work that we expect that optical emission
zone is sited towards the outer magnetosphere. Timing studies of the
size of the Crab pulse plateau indicates a restricted emission volume
($\approx$ 45 kms in lateral extent) (\cite{gold00}). This third point,
if consistent with the first two, probably points to emission coming
from a geometrically defined cusp along our line of sight akin to
treatment within the magnetosphere. More importantly the simple
relationships we have derived here indicate that there is no need to
invoke complex models for this high energy emission. Observed variations
in spectral index, pulse shape and polarization can be understood in
terms of geometrical factors rather differences in the production
mechanism. 

We note as well as that similar trends can be seen in $\gamma$ rays.
Figure 16 shows the correlation between the peak $\gamma$ emission as a
function of transverse field. This indicates a regression of the form\\

\noindent
Peak $\gamma$ Luminosity $\propto B_T^{0.89 \pm 0.15}$ \\ 

\noindent
significant at the 99.6 \% level, consistent with the observed steeper
distribution seen in the high energy $\gamma$-ray band (cf.~Table 2).

If we take our results and those of \cite{gold95} we can begin to
understand how the high energy emission process evolves with
canonically derived pulsar age. Goldoni et al compared the known
spectral indices and efficiencies in both the optical and $\gamma$-ray
regions. They noted that the spectral index flattened with age for the
$\gamma$-ray pulsars whilst the reverse was true for the optically
emitting systems. They also noted a similar trend reversal for the
efficiency with the $\gamma$-ray pulsars becoming more efficient with
age. We note (see the bottom panel in Figures 15 and 16) a similar
behaviour with the peak emission. If, as seems likely from the
temporal coincidence between the $\gamma$-ray and optical pulses, that
the source location is similar for both mechanisms. One explanation is
that we seem to have a position where by from the same electron
population there are two emission processes - expected if we have
curvature for the $\gamma$-ray photons and synchrotron for the optical
ones. It seems likely that the optical photon spectrum has been
further modified to produce the reversal in spectral index with
age. The region over which the scattering can occur would scale with
the size of the magnetosphere and hence with age. With the outer
magnetosphere magnetic fields for these pulsars being $< 10^6 $ G
electron cyclotron scattering is not an option. However synchrotron
self-absorption could explain the observed features. In essence we
would expect the most marked flattening to be for the Crab pulsar
where the outer field strengths are of order $10^6$ G and less so for
the slower and older systems.

These results (both optical and $\gamma$ rays) are consistent with a
model where the $\gamma$ and optical emission is coming from the last
open-field line at some constant fraction of the light cylinder. The
drop in efficiency with age for the production of optical photons
points towards an absorbing process in the outer magnetosphere.
Clearly more optical and $\gamma$-ray observations are needed to
confirm these trends.

\begin{figure}
\centerline{\psfig{file=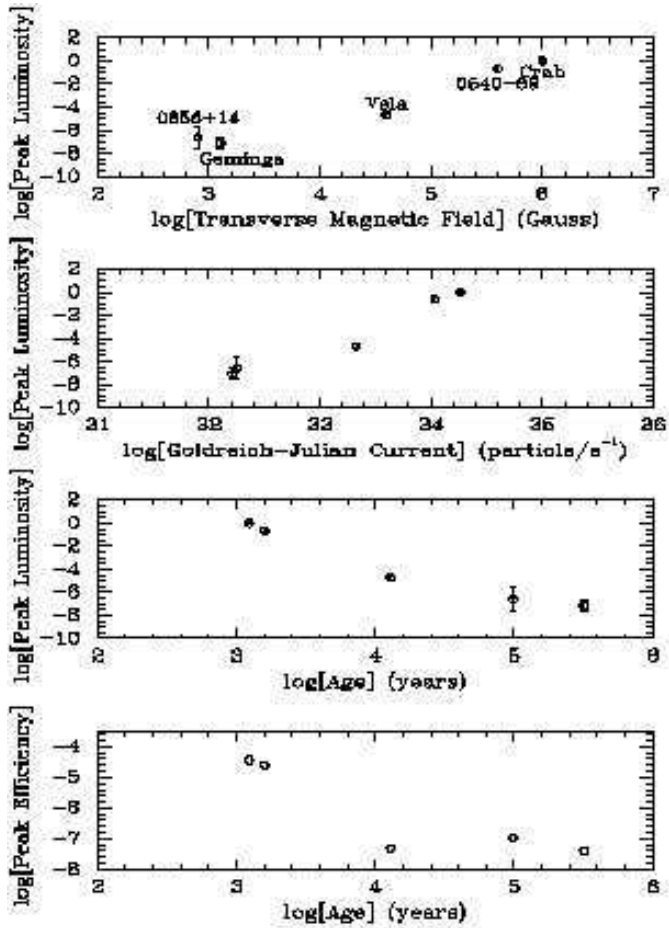,width=8.8cm,clip=} }
\caption{Peak optical luminosity vs Light Cylinder magnetic field strength, GJ Current and age. Also shown is the optical efficiency as a function of age.} 
\end{figure}

\begin{figure}
\centerline{\psfig{file=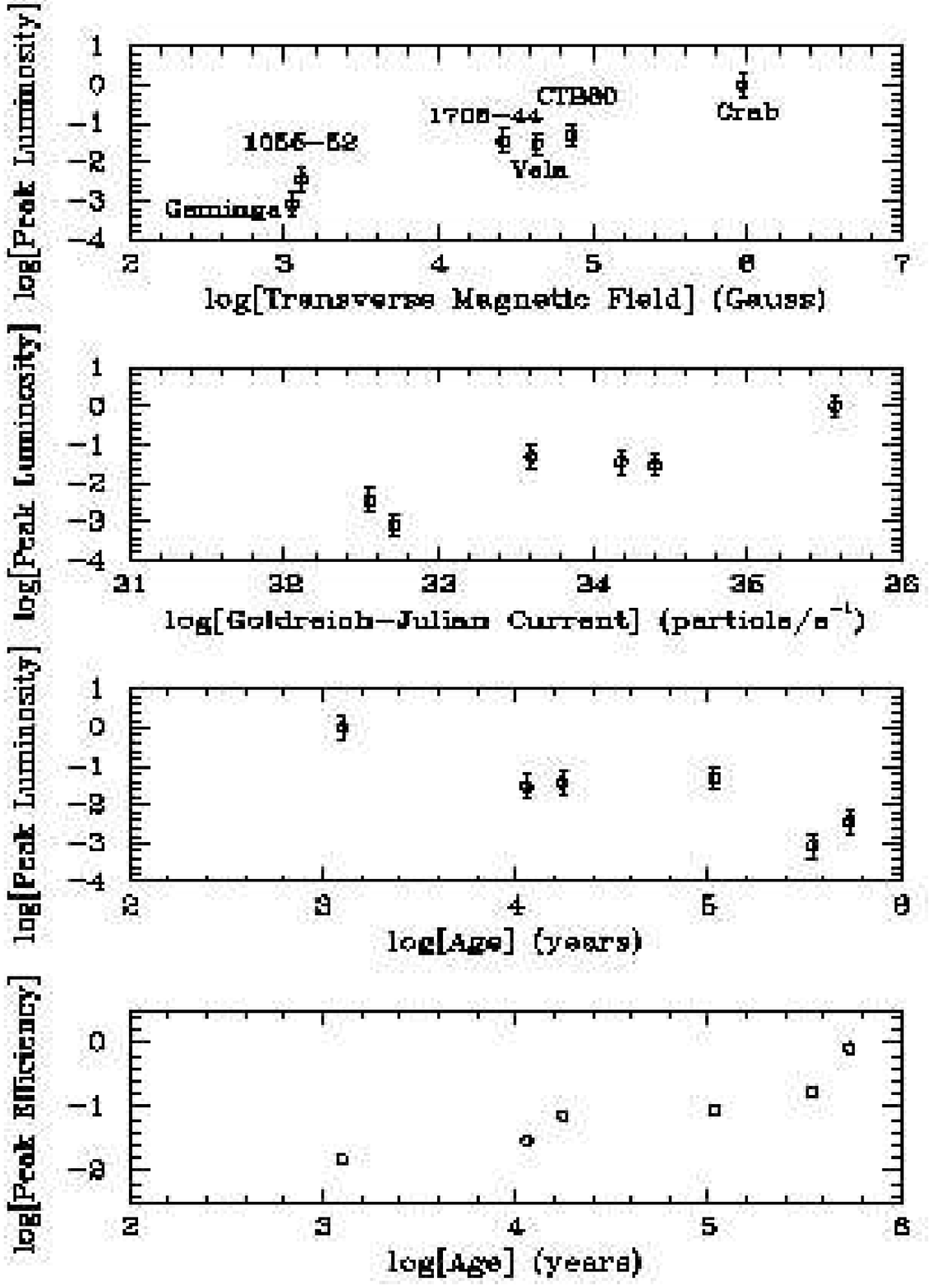,width=8.8cm,clip=} }
\caption{Outer field strength against peak $\gamma$ ray luminosity.  The
$\gamma$ ray peak luminosity has been inferred from \cite{fie98},
\cite{thom96}, \cite{thom99}}
\end{figure}

\begin{figure}
\centerline{\psfig{file=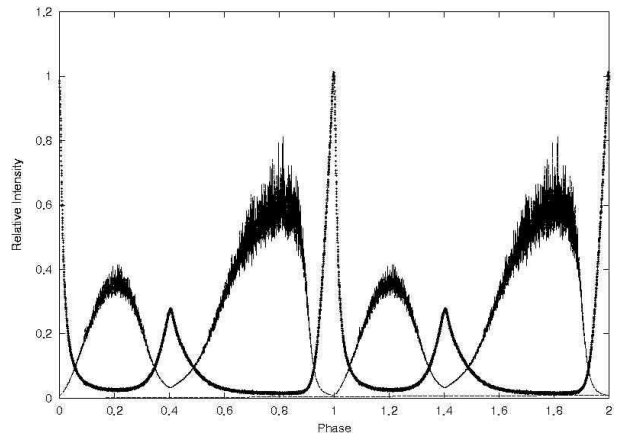,width=8cm,clip=} }
\caption{Crab light curve and inferred polarization fraction using Stokes parameters (Q/U) given 
by \cite{smith88} for the off pulse region and then scaled. The contribution from the main pulsar 
component is not shown.} 

\end{figure}

\section{Instrumental Considerations}

Table 3 summarizes the current range of detectors available for
studies of optical pulsars. To obtain the necessary background
measurements 2-d imaging detectors are required - these however
(e.g. the MAMA detector used in the TRIFFID photometer, \cite{but98})
have limited quantum efficiency. However the limited qe can be offset
by reduced effective sky aperture - the TRIFFID photometer uses
off-line sky apertures determined by the local seeing conditions.

\begin{table}[ht!!]
\caption{Current detector possibilities. The future should lie with the energy
sensitive detectors once the number of pixels can be increased}
\begin{small}
\begin{tabular}{lccc}
Detector & Time & Quantum & Number of \\
        & Resulution &  Efficiency & Pixels \\
\hline
CCDs       &  10 secs + &   70\% & millions \\
\hline
Fast CCDs \\
e.g. UltraCam &  100ms+  & 70\% & millions \\
\hline
APDs \\
     Optima    &     $\mu$ secs +  & 70\% & single \\
\hline
TRIFFID   &     $\mu$ secs +  & 70\% & single \\
Photocathode \\ based systems \\
        2-d - (e.g.) MAMA  &  $\mu$secs +  & $<$10\%    &      million \\
\hline
        Photomultipliers &   nsecs +  &    30\% max      & single \\
\hline
Energy Sensitive \\ Detectors \\
        TES     &          nsecs +     &      70\% max   & few \\
        STJ (SCAM)  &         nsecs +   &     70\%max    & few \\
\end{tabular}
\end{small}
\end{table}

CCD based detectors, e.g. ULTRACAM (\cite{dhil01}), suffer from
read-noise problems - which increase with decreasing period/time
resolution.  These will be effective for AXP studies but
'conventional' pulsars have too short a period. Techniques such as
on-chip phase locked image shifting and phase locked moving the
secondary mirror have limited application due to field crowding.

Avalanche Photodiodes - as used in OPTIMA (\cite{str01}) and TRIFFID - have good
quantum efficiency, timing, but, are only single pixel devices. In
both Optima and TRIFFID more than one APD is used for different
colours and sky background determination. Attempts are currently being
made to construct arrays of these items but there are serious
cross-talk problems. The next generation of energy sensitive detectors
(see for example, \cite{per99} \& \cite{rom99}) will give pulsar
observers almost all (with the exception of polarization) flux
parameters. However given that these devices are still in their
relative infancy there is still a place of APDs and 2-d photocathode
based systems.

\begin{figure}
\centerline{\psfig{file=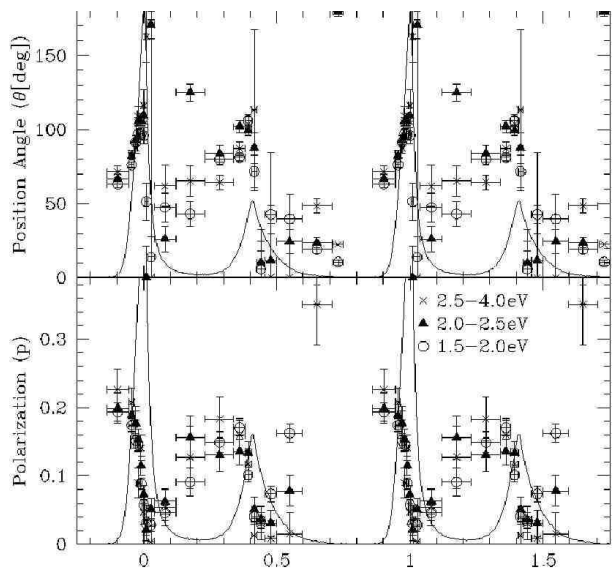,width=8.8cm,clip=} }

\caption{Crab polarization as a function of pulsar phase measured using the energy sensitive TES from \cite{rom01}}

\end{figure}

\section{Conclusion}

\begin{table*}
\caption{Main Characteristics of Optical Pulsars: B$_S$ \& B$_{LC}$ the
canonical surface and transverse magnetic field at the light cylinder
respectively; Opt. Lum and Peak Lum. refer to the optical luminosity at
the indicated distance in the B band.   }

\begin{small}

\begin{tabular}{lccccccccc}
Name & D      & P  & $\dot{P}$  &  $B_S$ & $B_{LC}$  & Int & Peak & Spectral Index & Cutoff \\
     &  (kpc) & (ms)   &  $10^{-14}$ s/s & log(G) & log (G)   & $\mu$Crab & $\mu$Crab & at 4500 \AA & \AA \\

Crab        & 2       & 33  & 40  & 12.6 & 6.1  & $10^6$ & $10^6$ & -0.11 & $>$20000\\
Vela        & 0.5     & 89  & 11  & 12.5 & 4.8  & 27 & 21  & 0.2 & 6500(?) \\
PSR0545-69  & 49      & 50  & 40  & 12.7 & 5.7  & $1.1~10^6$ & $1.4~10^5$  & 0.2 & $>$7000 \\
PSR0656+14  & 0.76(?) & 385 & 1.2 & 12.7 & 3.0  & 1.8 & 0.3  & 1.3 & $>$8000 \\
PSR0633+17     & 0.16    & 237 & 1.2 & 12.2 & 3.2  & 0.3 & 0.1  & 1.9 & $>$8000 \\
\end{tabular}
\end{small}
\end{table*}

\begin{table*}
\caption{Main Characteristics of $\gamma$-ray Pulsars: B$_S$ \& B$_{LC}$ the
canonical surface and transverse magnetic field at the light cylinder
respectively; $\gamma$ Lum and Peak Lum. refer to the $\gamma$ luminosity at
the indicated distance for E $>$ MeV.   }

\begin{small}

\begin{tabular}{lcccccccc}
Name & D      & P  & $\dot{P}$  &  $B_S$ & $B_{LC}$  & Int Lumin. & Peak Lumin. & Spectral Index  \\
     &  (kpc) & (ms)   &  $10^{-14}$ s/s & log(G) & log (G)   & Crab=1 & Crab=1 & \\

Crab           & 2       & 33  & 42.1 & 12.58 & 5.97  & 1 & 1 & 2.15 \\
Vela           & 0.5     & 89  & 12.5  & 12.53 & 4.64  & 0.0480 & 0.0299  &  1.70 \\
PSR1055-52     & 49      & 197  & 0.6  & 12.03 & 3.11  & 0.0124 & 0.0037  & 1.18 \\
PSR1706-44     & 2.4     & 102 & 9.3 & 12.49 & 4.42  & 0.1380 & 0.0368  & 1.72 \\
PSR0633+17     & 0.16    & 237 & 1.1 & 12.21 & 3.05  & 0.0019 & 0.0008  & 1.50 \\
PSR1951+32     & 2.5     & 40 & 0.6 & 11.69 & 4.86  & 0.0500 & 0.0187  & 1.74 \\
\end{tabular}
\end{small}

\end{table*}

The number of optically emitting pulsar which have been observed is
limited. The observed phenomenology points to emission in the outer
region of the magnetosphere - this seems more likely to come from an
Outer Gap type model rather than a Polar Cap. However the data is
sparse and no definitive statement can be made separating these models
on the basis of the optical emission alone.  Over the next few years
as with the advent of larger telescopes and more sensitive detectors, we
can confidently expect the number of optical detections of isolated
neutron stars to increase. PSR B1055-52 \& PSR B1951+32 are likely early
candidates. 

In the optical any potential thermal component should be separable
from the strongly pulsed magnetospheric emission, allowing for
reliable estimates of the neutron star radius to be measured with
consequent implications for equation of state models.  However before
this is done with any degree of confidence the nature of the unpulsed
emission must be established - is it a synchrotron knot feature or
more localized to the pulsar. Is it unique to the Crab pulsar?  Of
crucial importance in the future will be determination of the
low-energy cutoff and the polarization sweep through the optical pulse
(significantly modern instruments are capable of measuring the
polarization sweep for both PSR B0540-69 and Vela as well as the
Crab). Also of interest will be the shape of the pulse - in particular
the size of any plateau which scales as the size of the emission
zone. All of these parameters are measurable with existing
technologies and telescopes for these pulsars. The advent of new
detectors and larger telescopes should herald a period of renewed
interest in optical pulsar studies, however there are still important
questions that can be answered by observation with existing
telescope/instrument systems.

\section{Acknowledgments}

Enterprise Ireland is acknowledged for support under its Basic
Research Grant Scheme. Ray Butler and Padraig O'Connor are thanked 
for help during the preparation of this manuscript. The workshop 
organizers are thanked for their hospitality.


%% file: make_all.bbl
\begin{thebibliography}{}

\bibitem[Bignami et al (1988)]{big88} Bignami, G. F., Caraveo, P. A. \& Paul, J. A., 1988, A \& A, 202, L1 

\bibitem[Boyd et al (1995)]{boy95} Boyd, P. T. et al, 1995, ApJ, 448, 365 

\bibitem[Butler et al (1998)]{but98}Butler, R. F. \& Shearer, A. \& Redfern, R. M., Colhoun, M., O'Kane, P., Penny, A. J., Morris, P. W., Griffiths, W.K. \& Cullum, M., 1998, MNRAS, 296, 379

\bibitem[Cocke et al (1969)]{coc69}Cocke, W. J., Disney, M. J. \& Taylor, D. J., 1969, Nature, 221, 525 

\bibitem[Caraveo et al (1996)]{car96}Caraveo, P. A., Bignami, G. F., Mignani, R. \& Taff, L. G., 1996, A \& A, 120, 65 

\bibitem[Cheng et al (2000)]{che00}Cheng, K. S., Ruderman, M. \& Zhang, L., 2000,ApJ, 537, 964

\bibitem[Dhillon \& Marsh (2001)]{dhil01}Dhillon, V. \& Marsh, T, 2001, NewAR, 45, 91

\bibitem[Daugherty \& Harding (1996)]{dau96}Daugherty, J. K. \& Harding, A. K., 1996, ApJ, 458,278

\bibitem[Eikenberry \& Fazio (1997)]{eik97}Eikenberry, S. S. \& Fazio, G. G., 1997, ApJ, 476, 281 
\bibitem[Fierro et al (1998)]{fie98}Fierro, J. M., Michelson, P. F., Nolan, D. C. \& Thompson, D. J., 1998, ApJ, 494, 734

\bibitem[Gil et al (2000)]{gil00}Gil, J. A., Khechinashvili \& Melikidze, G. I., 2000, to be published in MNRAS

\bibitem[Golden \& Shearer(1999)]{gold99}Golden, A. \& Shearer, A., 1999, A \& A, 342, L5 

\bibitem[Golden et al (2000)]{gold00}Golden, A., Shearer, A. \& Beskin, G. M., 2000, ApJ, 535, 373 

\bibitem[Goldoni et al (1995)]{gold95}Goldoni, P., Musso, C., Caraveo, P. A. \& Bignami, G. F., 1995, \aa, 298, 535

\bibitem[Gouiffes et al  (1992)]{gou92} Gouiffes, C., Finley, J. P. \& Oegelman, H., 1992, ApJ, 394, 581

\bibitem[Gouiffes \& Ogelman (1996)]{gou96} Gouiffes, C. \& Ogelman, H., 1996 ASP Conf. Ser. 105: IAU Colloq. 160: Pulsars: Problems and Progress, 299


\bibitem[Halpern \& Tytler (1988)]{hal88} Halpern, J-P \& Tytler, D., 1988, ApJ, 330, 201 

\bibitem[Hester et al (1995)]{hes95} Hester, J. J. et al, 1995, ApJ, 448 240


\bibitem[Kaplan et al (2001)]{kap01}Kaplan, D. L., van Kerkwijk, M. H. \& Anderson, J., 2001, astro-ph/011174

\bibitem[Kawai et al (2002)]{kaw02}Kawai et al, 2002, these proceedings 

\bibitem[Nasuti et al (1996)]{nas96} Nasuti, F. P., Mignani, R., Caraveo, P. A. \& Bignami, G. F., 1996, A\&A, 314, 849 

\bibitem[Martin et al (1998)]{mar98}Martin, C, Halpern, J.P. \& Schiminovich, D., 1998, ApJ, 494, L211 

\bibitem[Middleditch \& Pennypacker(1985)]{mid85}Middleditch, J. \& Pennypacker, C., 1985, Nature, 313, 659 

\bibitem[Middleditch et al (1987)]{mid87}Middleditch, J., Pennypacker, C. \& Burns, M. S., 1987, ApJ, 315, 142

\bibitem[Mignani et al (1998)]{mig98}Mignani, R. P., Caraveo, P. A., \& Bignami, G. F., 1998, \aa, 332, L37 

\bibitem[Nasuti et al (1996)]{nas97} Nasuti, F. P., Mignani, R., Caraveo, P. A. \& Bignami, G. F., 1996, A\&A, 323, 839 


\bibitem[Pacini (1971)]{pac71}Pacini, F., 1971, ApJ, 163,17 

\bibitem[Pacini \& Salvati (1983)]{pac83}Pacini, F.  \& Salvati, M., 1983, ApJ, 274, 369 
\bibitem[Pacini \& Salvati (1987)]{pac87}Pacini, F.  \& Salvati, M., 1987, ApJ, 321, 445 

\bibitem[Pavlov et al (1997)]{pav97} Pavlov, G. G., Welty, A. D. \& Cordova, F. A., 1997, ApJ, 489, L75

\bibitem[Perryman et al (1999)]{per99}Perryman, M. A. C., Favata, F., Peacock, A., Rando, N. \& Taylor, B. G., 1999, A \& A, 346, 30 

\bibitem[Peterson et al (1978)]{pet78} Peterson, B. A., Murdin, P., Wallace, P., Manchester, R. N., Penny, A. J., Jorden, A., Hartley, K. F. \& King, D, 1978, Nature, 276, 475

\bibitem[Romani \& Yadigaroglu (1995)]{rom95}Romani, R. W., \& Yadigaroglu, I.-A., 1995, ApJ, 438, 314  

\bibitem[Romani et al (2001)]{rom01}Romani, R. W., Millar, A. J., Cabrera, B., Nam, S. W. \& Yadigaroglu, I.-A., 2001, astro-ph 0108240  

\bibitem[Romani et al (1999)]{rom99}Romani, R. W., Miller, A. J., Cabera, B. \& Figueroa-Feliciano, E., 1999, ApJ, 521, L151 

\bibitem[Shearer et al (1997)]{shear97}Shearer, A., Redfern, R.  M., Gorman, G., Butler, Golden, A., R., O'Kane, P., Golden, A., Beskin, G.  M., Neizvestny, S.  I., Neustroev, V.  V., Plokhotnichenko, V.  L.  \& Cullum, M., ApJ, 1997,
 487, L181

\bibitem[Shearer et al (1998)]{shear98}Shearer, A., Harfst,
 S., Redfern, R.  M., Butler, R., O'Kane, P., Beskin, G.  M.,
 Neizvestny, S.  I., Neustroev, V.  V., Plokhotnichenko, V.  L.  \&
 Cullum, M., 1998, A \& A, 335, L21 

\bibitem[Shearer \& Golden
 (2001)]{shear01} Shearer, A. \& Golden, A., 2001, ApJ, xxx, yyy 

\bibitem[Shearer et al (2002a)]{shear02a} Shearer, A., O Connor, P. \& Golden, A., Ryan, O., \& Redfern, M., in preparation 

\bibitem[Shearer et al (2002b)]{shear02b} Shearer, A., Butler, R. F. \& Golden, A., in preparation 

\bibitem[Smith et al (1988)]{smith88} Smith, F. G.,  Jones, D. H. P., Dick, J. S. P. \& Pike, C. D., 1988, MNRAS, 233, 305
 
\bibitem[Sollerman et al (2000)]{sol00} Sollerman, J., Lundqvist, P., Lindler, D., Chevalier, R. A., Fransson, C., Gull, T. R., Pun, C. S. J. \& Sonneborn, G, 2000, ApJ, 537, 861

\bibitem[Sollerman \& Flyckt (2002)]{sol02} Sollerman, J. \& Flyckt, V., 2002, ESO Messenger, 107, 32

\bibitem[Straubmeier et al (2001)]{str01} Straubmeier, C., Kanbach, G. \& Schrey, F., 2001, Exp Ast, 11, 157

\bibitem[Thompson et al (1996)]{thom96}Thompson, D. J., 1996, ApJ,
 465, 385 \bibitem[Thompson et al (1999)]{thom99}Thompson, D. J.,
 1999, ApJ, 516, 297

\bibitem[Wallace et al (1977)]{wall77}Wallace,
 P. T. et al. 1977, Nature, 266, 692 

\end{thebibliography}
